\documentclass[aps,pra,twocolumn,showpacs,letterpaper,superscriptaddress]{revtex4-1}

\usepackage[colorlinks=true, citecolor=blue, linkcolor=blue, urlcolor=blue]{hyperref}

\usepackage{graphicx,dcolumn,longtable,epsfig}
\usepackage[usenames]{color}
\usepackage{amssymb}
\usepackage{amsmath}
\usepackage{epstopdf}
\usepackage{bm}
\usepackage{footnote}
\usepackage{float}
\usepackage{subfigure}
\usepackage{color}
\usepackage{ulem}

\input epsf.tex

\def\bea{\begin{eqnarray}}
\def\eea{\end{eqnarray}}
\def\be{\begin{equation}}
\def\ee{\end{equation}}

\begin{document}

\title{Equilibrium phases of dipolar lattice bosons in the presence of random diagonal disorder}
\author{ C. Zhang }
\affiliation{Homer L. Dodge Department of Physics and Astronomy,
The University of Oklahoma, Norman, Oklahoma ,73019, USA}
\author{ A.~Safavi-Naini}
\affiliation{JILA , NIST and Department of Physics, University of Colorado, 440 UCB, Boulder, CO 80309, USA}
\author{B. Capogrosso-Sansone}
\affiliation{Department of Physics,
Clark University, Worcester, Massachusetts, 01610, USA}

\begin{abstract}
Ultracold gases offer an unprecedented opportunity to engineer disorder and interactions in a controlled manner. In an effort to understand the interplay between disorder, dipolar interaction and quantum degeneracy, we study two-dimensional hard-core dipolar lattice bosons in the presence of on-site bound disorder. Our results are based on large-scale path-integral quantum Monte Carlo simulations by the Worm algorithm. We study the ground state phase diagram at fixed half-integer filling factor for which the clean system is either a superfluid at lower dipolar interaction strength or a checkerboard solid at larger dipolar interaction strength. We find that, even for weak dipolar interaction, superfluidity is destroyed in favor of a Bose glass at relatively low disorder strength. Interestingly, in the presence of disorder, superfluidity persists for values of dipolar interaction strength for which the clean system is a checkerboard solid. At fixed disorder strength, as the dipolar interaction is increased, superfluidity is destroyed in favor of a Bose glass. As the interaction is further increased, the system eventually develops extended checkerboard patterns in the density distribution. Due to the presence of disorder, though, grain boundaries and defects, responsible for a finite residual compressibility, are present in the density distribution. Finally, we study the robustness of the superfluid phase against thermal fluctuations.
\end{abstract}

\maketitle

\section{Introduction}
Since their first theoretical investigation three decades ago~\cite{Ma:1986ge, Giamarchi:1987du, Giamarchi:1988dv, Fisher1}, many-body bosonic systems in the presence of disorder have attracted a great deal of attention both experimentally and theoretically~\cite{Schwartz:2007wi, Hu:2008jy, Crowell:1997gm, Goldman:1998jy, Schneider:2012id, vanderZant:1996cf, Inguscio2008, Inguscio2010, DeMarco2009PRL, DeMarco2010Nature, Schneble2011, DErrico:2014gc, Rapsch:1999bw, Ceperley1991, Pollet2009BG, Pollet2009PRB, Soyler:2011ik, Ceperley2011, Zhang:2015it, Niederle:2013jy, Pollet:2009dg, Gurarie:2009it}. While Anderson localization~\cite{Schwartz:2007wi, Hu:2008jy, Inguscio2008} and classical trapping of non-interacting bosons~\cite{RobertdeSaintVincent:2010gr} are well understood, the interplay between disorder, interaction and quantum degeneracy in strongly-correlated bosonic systems may give rise to new fascinating phenomena. A certain degree of disorder is ubiquitous in condensed matter systems, but a thorough understanding of these systems is hindered by poor control over the nature of disorder and competing interactions. Ultracold gases, on the other hand, offer an unprecedented level of control over interactions and disorder. Specifically, interactions and disorder can be tuned independently. Experimentally, the most common way to generate disorder is via optical speckle fields~\cite{CreatDisorder1,DeMarco2009PRL,AAspect2012}. Other possibilities to engineer disorder include quasi-periodic potentials generated by non-commensurate bichromatic optical lattices~\cite{CreatDisorder2} , the introduction of impurity atoms to the system~\cite{Schneble2011}, and holographic techniques which produce point-like disorder~\cite{Morong:2015gh}. 

A paradigmatic example is that of lattice bosons described by the Bose-Hubbard model in the presence of on-site, bound disorder, where it was proven analytically and confirmed numerically that the gapless Bose glass (BG) phase, characterized by finite compressibility and absence of off-diagonal long-range order, always intervenes between the Mott-insulator and superfluid (SF) state~\cite{Pollet:2009dg, Gurarie:2009it}.
Moreover, numerical studies at commensurate and incommensurate filling factor have shown that large enough disorder always destroys superfluidity in favor of the BG. In the weakly-interacting limit, interactions compete with Anderson localization thus enhancing superfluidity~\cite{Gurarie:2009it, Lugan:2007fy}. 
This results in sizable disorder strength needed in order to destroy superfluidity. On the other hand, in the strongly-interacting limit, interactions suppress superfluidity and, at integer filling, eventually completely destroy it, leaving the system either in the Mott insulator phase at lower disorder strength or in a BG phase at larger disorder strength~\cite{Gurarie:2009it, Soyler:2011ik}.

In this paper, we consider dipolar lattice bosons in two-dimensions, as described by the extended Bose-Hubbard model, in the presence of on-site bound disorder. Dipolar lattice systems are now accessible experimentally. They can be realized with polar molecules~\cite{Yan:2013fn, Hazzard:2014bx}, atoms with large magnetc moments~\cite{DePaz:2013ff, Lu:2011hl}
, and Rydberg atoms~\cite{Saffman:2010ky, Lw:2012ct, Gunter:2013fv}. Unlike single-component atomic systems purely interacting via Van-der-Waals interactions, dipolar systems interacting via the long-ranged and anisotropic dipolar interaction can realize novel superfluid, solid, and topological phases~\cite{Potter:2010hk, CapogrossoSansone:2010em, Zhang:2015is, Gorshkov:2011ct, Gong:2016hh, Gong:2016df, Yao:2012fn}.  
While in-depth theoretical studies of dipolar lattice bosons in presence of disorder are still lacking, as suggested in~\cite{Deng:2013db}, many of these phases may not be robust in presence of disorder. 

In this paper we study two-dimensional hard-core dipolar bosons trapped in a square optical lattice and in the presence of random on-site disorder and use large-scale quantum Monte Carlo simulations by the Worm algorithm~\cite{Prokofev:1998gz} to study the robustness of the equilibrium phases to disorder. In particular we show that the interplay of the long-range interactions and the on-site disorder leads to the suppression of the checkerboard (CB) order, while enhancing the SF order, and stabilizing a BG phase.

\section{System Hamiltonian} 
The system is comprised of hard-core, dipolar lattice bosons in a 2d square lattice, with the dipole moments aligned perpendicular to the lattice by an external static electric field so that the dipolar interaction is purely repulsive. At half-integer filling and in the absence of disorder, the system is either in a SF state at lower dipolar interaction strength or in a checkerboard (CB) solid phase at larger dipolar interaction strength~\cite{CapogrossoSansone:2010em}. In the presence of on-site disorder, the system is described by the Hamiltonian:
\begin{equation}
\nonumber H=-J\sum_{\langle i\,j\rangle }a_i^\dagger a_j+V\sum_{i<j}{\frac{n_{i}n_{j}}{r_{ij}^{3}}}- \sum_i (\varepsilon_{i}-\mu)n_i \;\; ,
\label{equation1}
\end{equation}
where the first term is the kinetic energy characterized by hopping amplitude $J$. Here $\langle \cdots \rangle$ denotes nearest neighboring sites, $a_i^\dagger$ ($a_i$) are the bosonic creation (annihilation) operators satisfying the usual commutation relations and the hard-core constraint $a_i^\dagger a_i^\dagger=0$. The second term is the purely repulsive dipolar interaction characterized by strength $V=d^2/a^3$, $d$ is the induced dipole moment and $a$ is the lattice spacing,  $r_{ij}$ is the relative distance (measured in units of $a$) between site $i$ and site $j$, $n_i=a_i^{\dagger}a_i$ is the particle number operator. The third term is the chemical potential term with chemical potential $\mu$ shifted by the on site random disorder potential $\varepsilon_{i}$, where $\varepsilon_{i}$ is uniformly distributed within the range $[\Delta,-\Delta]$. We set our unit of energy and length to be the hopping amplitude $J$ and the lattice spacing $a$, respectively.

\section{Ground state phase diagram}\label{sec:phasediag}
\begin{figure}[h]
\includegraphics[trim=0cm 0cm 3cm 2cm, clip=true, width=0.55\textwidth]{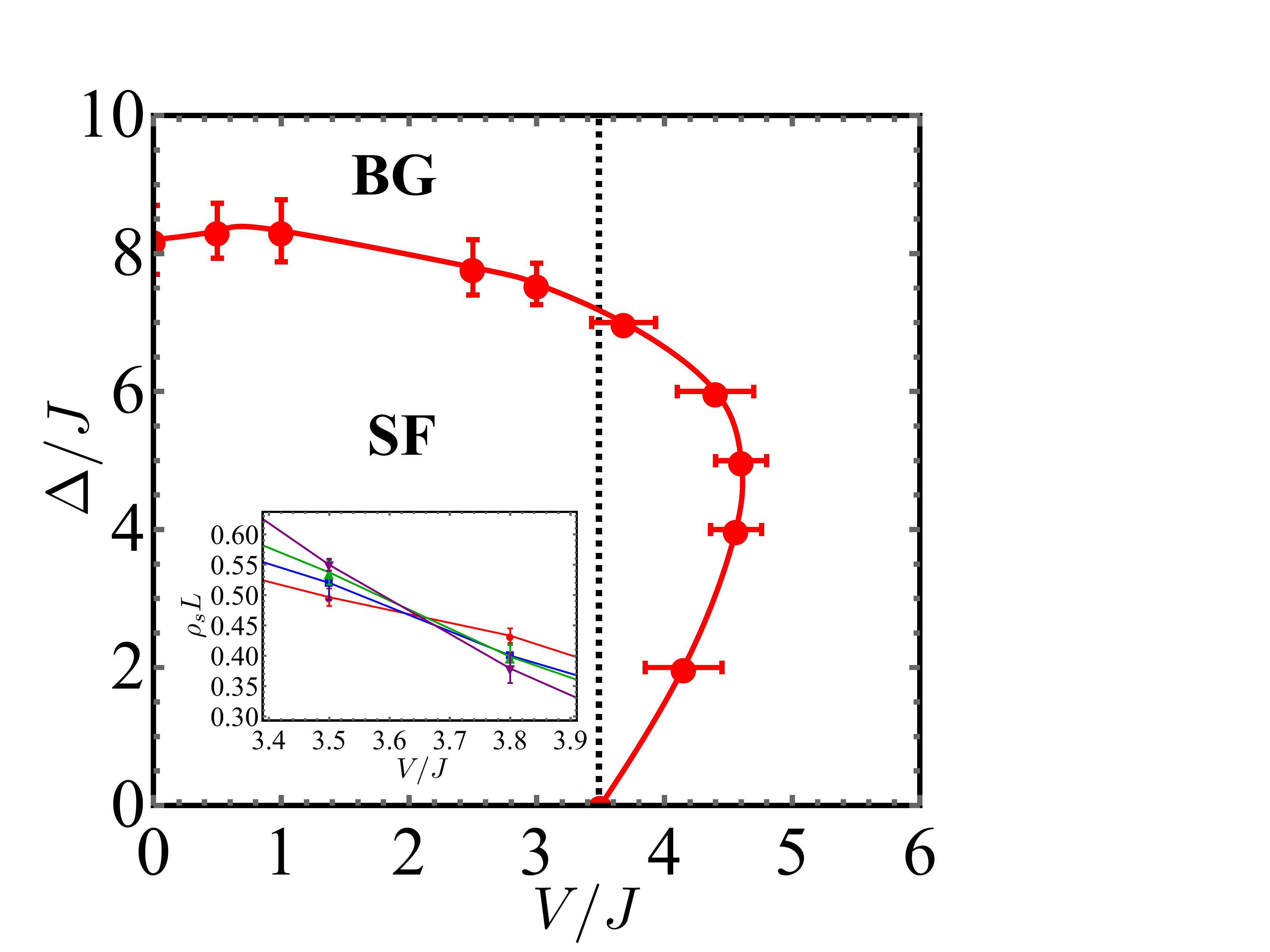}
\caption{(Color online) Phase diagram of the system described by Eq.~\ref{equation1} at filling factor $n=0.5$. The horizontal and vertical axes are the dipole-dipole repulsive interaction strength $V/J$ and the disorder strength $\Delta/J$, respectively. Solid red circles represent superfluid to insulator transition points as determined by standard finite size scaling. The solid line is a guide to the eye. The black dotted line is the interaction strength at which superfluidity disappears in favor of a checkerboard solid in the clean system. The inset shows finite size scaling of the superfluid stiffness $\rho_s$. We plot $\rho_s L$ vs. $V/J$ at $\Delta/J=7$ for system sizes $L=12$, 16, 20 and 24 (red circles, blue squares, green up triangles and purple down triangles, respectively). The transition point corresponds to the value of $V/J$ where curves referring to different system sizes cross. Here $V_c/J=3.68\pm 0.25$. Error bars are within the symbols if not visible in the plots.}
\label{dipolarphase}
\end{figure}

In this section, we present our numerical results for the ground state phase diagram of model~\ref{equation1} at fixed filling factor $n=0.5$, as shown in Fig.~\ref{dipolarphase}. The horizontal and vertical axes are the dipole-dipole interaction strength $V/J$ and the disorder strength $\Delta/J$, respectively. Red circles represent SF-insulator transition points. The red solid line is a guide to the eye. At zero disorder, the system is either in the SF phase for $V/J<3.5$, or the CB phase (dotted line)~\cite{CapogrossoSansone:2010em}. The SF phase is characterized by finite superfluid stiffness $\rho_s$ which can be determined from the statistics of winding numbers in space. Specifically, $\rho_{s}=\langle \mathbf{W}^2\rangle/dL^{d-2}\beta$, where $\mathbf{W}^2 =W_x^2+W_y^2$, $W_{x,y}$ being the winding number in spatial directions $x$ and $y$, $d=2$ is the spatial dimension, and $\beta$ is the inverse temperature~\cite{Ceperley:1989hb}. The CB solid possesses diagonal long-range order and is characterized by a finite value of structure factor $S(\mathbf{k})=\sum_{\mathbf{r},\mathbf{r'}} \exp{[i \mathbf{k} (\mathbf{r}-\mathbf{r'})]\langle n_{\mathbf{r}}n_{\mathbf{r'}}\rangle}/N$, here $\mathbf{k}$ is the reciprocal lattice vector. For CB solid, $\mathbf{k}=(\pi, \pi)$. 

For $V/J<3.5$, the SF is destroyed in favor of a BG phase for disorder strength $\Delta/J >7\div 8$. This is in contrast to what was found for the Bose-Hubbard model at unit filling and in the limit of weak interactions where sizable disorder strength $\Delta/J \sim 75$ is needed in order to destroy superfluidity~\cite{Soyler:2011ik}. This can be easily understood as follows. In the weak interaction limit, interactions compete with Anderson localization resulting in an enhancement of superfluidity. Here, instead, the hard-core nature of bosons suppresses superfluidity even at weak dipolar interaction. As a matter of fact, for $V/J<3.5$ superfluidity is destroyed in favor of the BG at a nearly constant value of disorder strength $\Delta/J\sim 7\div 8$, that is, the critical disorder strength is nearly independent of the dipolar interaction strength. 
Furthermore, at finite disorder strength $\Delta/J\lesssim 7$, the SF phase persists for dipolar interaction $V/J>3.5$, whereas the clean system featured a CB solid at this point. We attribute this to the competition between dipolar interaction which favors the CB order and disorder which tends to destroy it, resulting in a superfluid phase persisting beyond $V/J\sim 3.5$. Superfluidity is eventually completely destroyed for $V/J \gtrsim 4.6$.

With the exception of the transition point corresponding to $\Delta/J=2.0$, the SF to insulator transition appears to be of second order as demonstrated by the finite-size scaling of the superfluid stiffness $\rho_s$ where we have used the dynamical critical exponent $z=1$ (see inset of Fig.~\ref{dipolarphase}). For each $V/J$ and $\Delta/J$, we average over 500-1000 realizations of disorder, with disorder strength uniformly and randomly distributed within the interval $[-\Delta,\Delta]$. We have used system sizes L=12, 16, 20, 24. The transition point corresponds to the value of $V/J$ where curves referring to different system sizes cross ($V_c/J=3.68\pm 0.25$ in this example). We note that even for $V/J>3.5$ the SF disappears in favor of the BG phase rather than a first-order phase transition to the diagonally ordered CB phase. This is further confirmed by the observation of finite compressibility $\kappa=\beta (\langle n^2\rangle -\langle n \rangle ^2 )$, measuring density fluctuations, across the transition (see below).
\begin{figure}[h]
\includegraphics[trim=0cm 5cm 0cm 10cm, clip=true, width=0.5\textwidth]{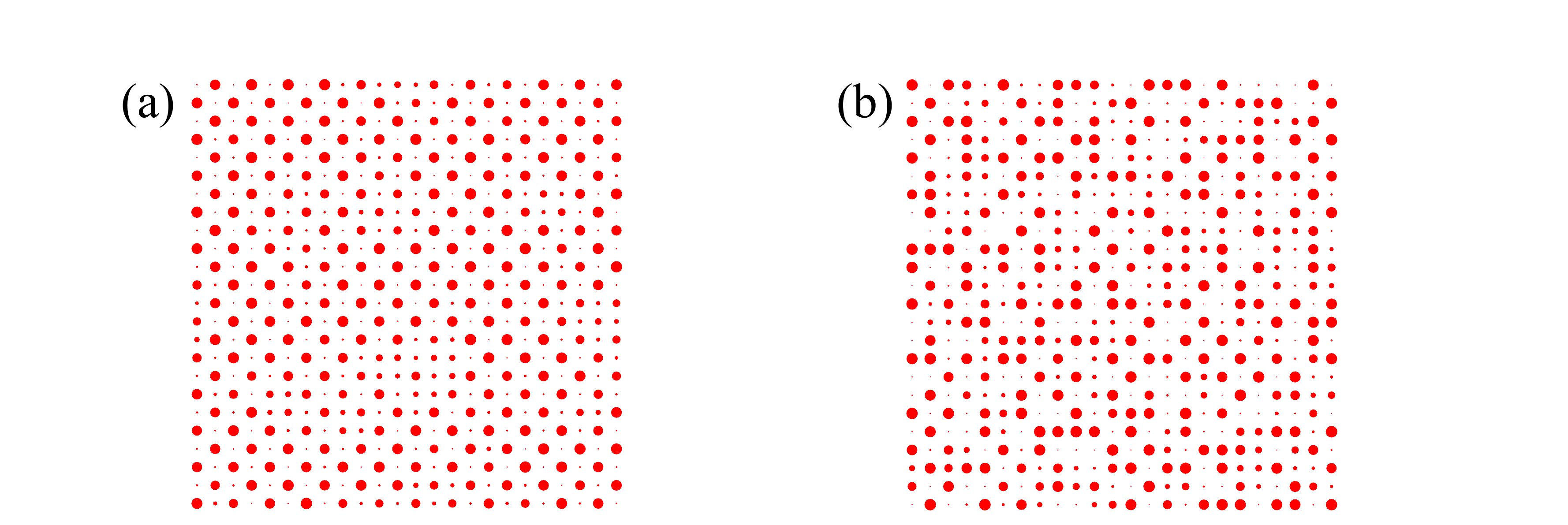}
\caption{(Color online) Imaginary-time average of the density distribution for a given Monte Carlo configuration and a specific disorder realization at (a) $V/J=4.2$, $\Delta/J=2$, (b) $V/J=4.0$, $\Delta/J=7$.}
\label{density_map1}
\end{figure}

The computational cost of our simulations limit us to system sizes of $L=100$. As such, we are unable to perform finite size scaling at fixed $\Delta/J=2$, and estimate the transition point to be within the range delimited by the interaction value where we start seeing finite size effects and the interaction value at which $\rho_s$ for the largest L (L=100) goes to zero.  Unlike what we observed for larger values of disorder strength, at $\Delta/J=2$ the superfluid phase is destroyed in favor of an insulating phase where the CB pattern is stabilized everywhere in the lattice apart from small superfluid regions. This can be seen in Figure~\ref{density_map1} (a) where we plot the imaginary-time average of the density distribution for a given Monte Carlo configuration and for a specific disorder realization at $V/J=4.2$, $\Delta/J=2$. Here, the radius of each circle is proportional to the density at that site. For comparison, in Figure~\ref{density_map1} (b), we show the density map at $V/J=4.0$, $\Delta/J=7$ where the CB patterns are clearly absent. While we observe coexistence of the SF and CB phases our present results do not allow us to determine the nature of the SF-insulator transition. 

\begin{figure}[h]
\includegraphics[trim=0cm 5.2cm 0cm 6cm, clip=true, width=0.5\textwidth]{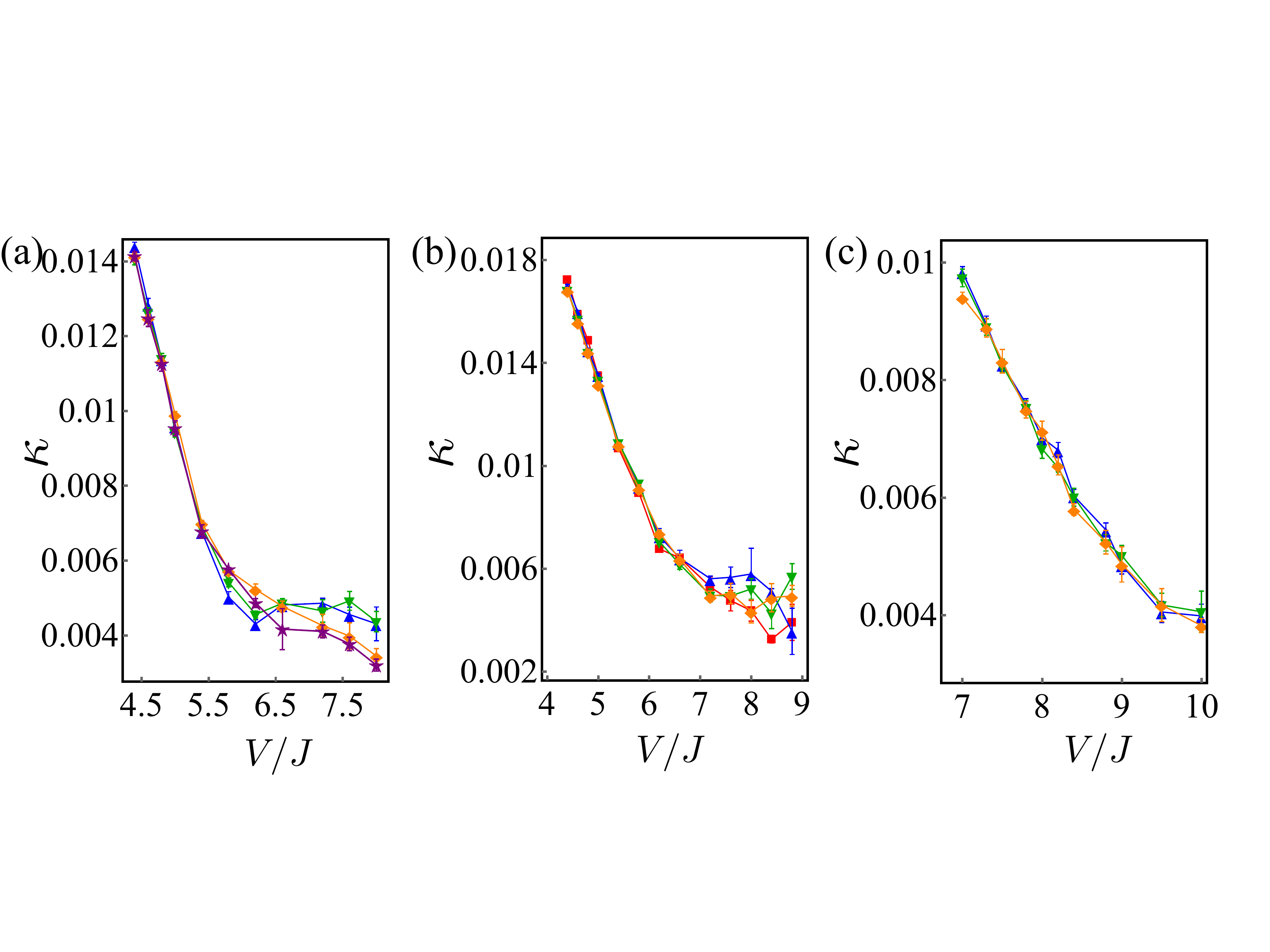}
\caption{(Color online) Compressibility $\kappa$ as a function of $V/J$ for system sizes L=12, 16, 20, 24 and 28 (red squares, blue up triangles, green down triangles, orange diamonds and purple stars, respectively) at fixed disorder strength $\Delta/J=4$ (a), 5 (b) and 7 (c). Compressibility remains finite beyond the superfluid to insulator transition confirming that the superfluid phase disappears in favor of a Bose glass. At large enough interaction, $\kappa$ seems to plateau. Upon approaching the plateau, extended CB patterns appear in the density distribution (see text for details). Error bars are within the symbols if not shown in the figures.}
\label{com}
\end{figure}

\begin{figure}[h]
\includegraphics[trim=0cm 5cm 0cm 10cm, clip=true, width=0.5\textwidth]{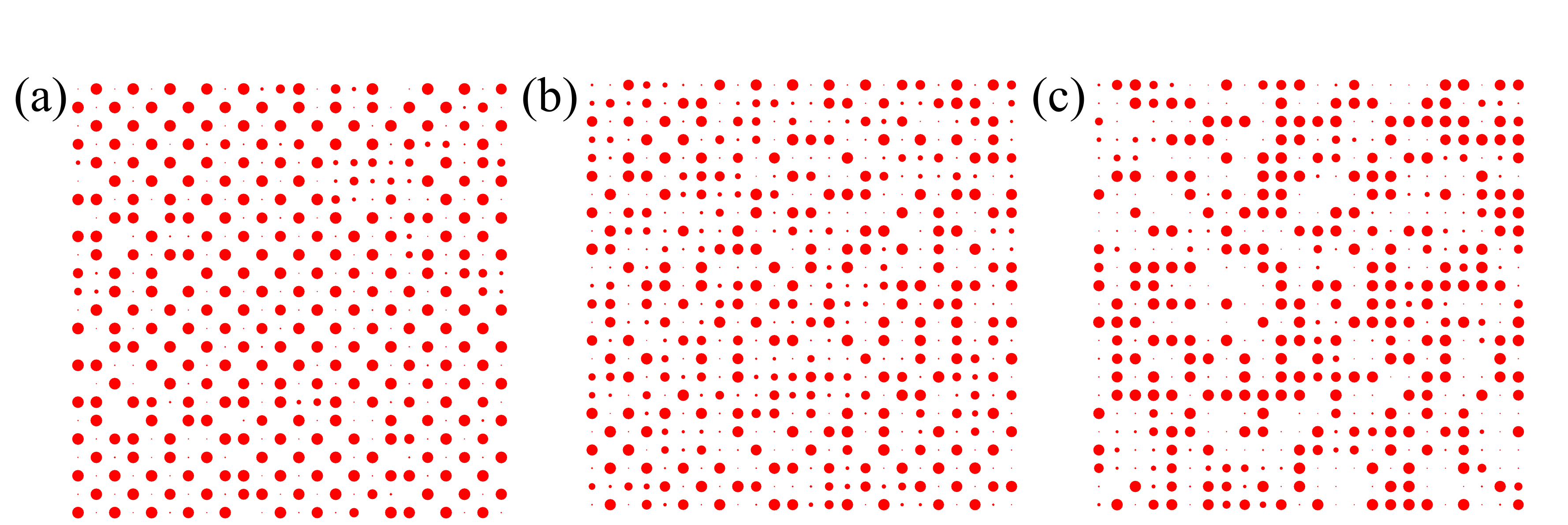}
\caption{(Color online) Imaginary-time average of the density distribution for a given Monte Carlo configuration and a specific disorder realization at (a) $V/J=10.0$, $\Delta/J=7.0$, (b) $V/J=4.5$, $\Delta/J=7.0$ and (c) $V/J=1.0$, $\Delta/J=9.5$. The radius of each circle is proportional to the density at that site. Defects and grain boundaries within the checkerboard order are present in (a). They contribute to a small residual compressibility.}
\label{density_map}
\end{figure}

As the dipolar interaction strength is increased at fixed disorder strength, one expects the BG to eventually disappear in favor of the CB solid, the latter being characterized by diagonal long range order and zero compressibility. We have performed scans in interaction strength at fixed $\Delta/J$. Figure~\ref{com} shows compressibility $\kappa$ as a function of $V/J$ at fixed $\Delta/J=4$, 5 and 7 (Figure~\ref{com} (a), (b), (c), respectively) for system sizes ranging from $L=12$ to 28. 
We identify the SF to insulator transition points for $\Delta/J=4$, 5 and 7 to be at $V/J=4.56\pm 0.2$, $4.6\pm 0.2$ and $3.68\pm 0.25$ respectively. Clearly, the compressibility remains finite beyond the transition confirming that the SF disappears in favor of a  BG, rather than the CB solid. Notice that at a given interaction strength, $\kappa$ is larger for larger disorder strength. Indeed, in this region of the parameter space, disorder competes with interactions which stabilize the incompressible CB solid. The larger the disorder strength, the larger the interaction strength needed in order to observe extended CB patterns in the density distribution.  As expected, in all cases, compressibility decreases as interaction increases. At large enough interaction, $\kappa$ seems to plateau. The larger the $\Delta/J$, the larger the interaction at which the plateau is observed. Upon approaching the plateau, extended CB patterns appear in the density distribution. However, due to the presence of disorder, defects and grain boundaries persist even at increasing interaction strengths and contribute to a small residual compressibility. In Fig.~\ref{density_map} (a) we plot the imaginary-time average of the density distribution for a given Monte Carlo configuration and for a specific disorder realization at $V/J=10.0$, $\Delta/J=7.0$. For comparison, we show the density maps in the BG phase at $V/J=4.5$, $\Delta/J=7.0$ (Fig.~\ref{density_map} (b)) and at $V/J=1.0$, $\Delta/J=9.5$ (Fig.~\ref{density_map} (c)). For $V/J>3.5$, a specific disorder realization  featuring regions that locally mimic a clean system allows the dipolar interactions to stabilize the CB phase in those regions. Such islands of CB phase are observed in Fig.~\ref{density_map} (b), but are absent from Fig.~\ref{density_map} (c) where the strength of interactions, $V/J<3.5$ is below the critical strength required to stabilize the CB phase in the clean system. 
We note that the disappearance of the BG in favor of a CB solid happens over a range of increasing  $V/J$ values at a fixed $\Delta/J$. As  $V/J$ increases, islands of CB within the BG phase grow larger and eventually the system settles into a CB with localized defects.

\section{Finite temperature results}

\begin{figure}[h]
\includegraphics[trim=2cm 1cm 2cm 1cm, clip=true, width=0.48\textwidth]{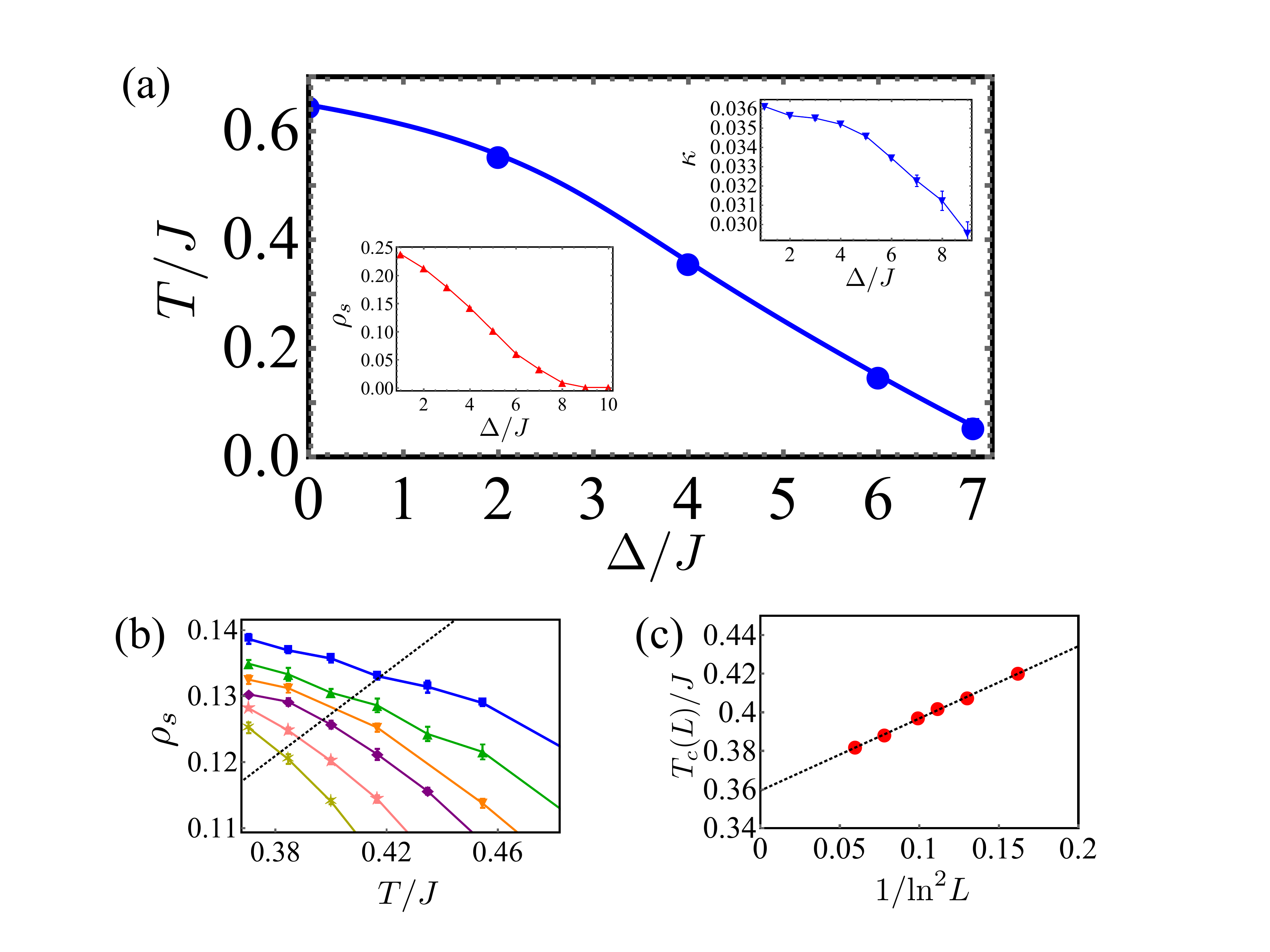}
\caption{(Color online) All plots refer to $V/J=2$. (a) Main plot: critical temperature $T_c/J$ for disappearance of superfluidity via a KT transition as a function of $\Delta/J$. Bottom-left inset: superfluid stiffness $\rho_s$ as a function of $\Delta/J$ for system size L=24. Top-right inset: Compressibility $\kappa$ as a function of $\Delta/J$ for system size L=24. (b) Superfluidity $\rho_s$ as a function of $T/J$ for L=12, 16, 20, 24, 36 and 60 (blue squares, green up triangles, purple down triangles, orange diamonds, pink stars and yellow asterisks, respectively), at $\Delta/J=4$. Dotted line corresponds to $\rho_s=T/\pi$. Its intersection points with each $\rho_s$-curve give `critical' temperatures $T_c(L)/J$ for a finite system. (c) $T_c(L)/J$ vs. $1/\text{ln}^2L$. Error bars are within symbol size if not visible in the plots.}
\label{FiniteT}
\end{figure}

In this section we investigate the robustness of the SF state against thermal fluctuations. Upon increasing the temperature, thermal fluctuations destroy superfluidity via a Kosterlitz-Thouless (KT) transition~\cite{Kosterlitz:1973fc}. In the following, we fix interaction strength to $V/J=2$ and scan over disorder strength to find the critical temperature at which the SF disappears. 
Figure~\ref{FiniteT} (a) shows the critical temperature $T_c/J$ as a function of $\Delta/J$. The top-right inset shows the ground state compressibility $\kappa$ as a function of disorder strength $\Delta/J$ for $L=24$. The bottom-left inset shows the ground state superfluid stiffness $\rho_s$ vs. $\Delta/J$ for $L=24$. While compressibility does not change significantly,  $\rho_s$ is suppressed as disorder strength is increased. As a result, the corresponding critical temperature decreases as a function of $\Delta/J$, as seen in the main plot. The critical temperature is found using standard finite size scaling. In the thermodynamic limit, a universal jump is observed at a critical temperature given by $\rho_{s}(T_c)=2m k_B T_c/\pi \hbar^2$. In a finite size system this jump is smeared out as seen in Figure~\ref{FiniteT} (b) which shows $\rho_{s}$ as a function of $T/J$ at $\Delta/J=4$ for system sizes $L=12$, 16, 20, 24, 36 and 60. The dotted line in Figure~\ref{FiniteT} (b) corresponds to $\rho_s=T/\pi$ and its intersection points $T_c(L)/J$ with each $\rho_s$-vs.-$T/J$-curve are used to find $T_c/J$ as shown in Figure~\ref{FiniteT} (c). Here, we plot $T_c(L)/J$ vs. $1/\text{ln}^2L$ and find the critical temperature at $\Delta/J=4$ to be $T_c/J \sim 0.36$.   Finally we note that the critical interactions strengths corresponding to the stabilization of each equilibrium phase are achievable in current experiments. However further improvements will be needed, both in cooling and loading phases, to achieve the required lattice gas temperatures.

\section{Conclusion}
We have studied hard-core bosons trapped in a square lattice, interacting via purely repulsive dipolar interaction and in the presence of on-site bound disorder. Our results are based on large-scale path-integral quantum Monte Carlo simulations by the Worm algorithm. We have presented the ground state phase diagram at fixed half-integer filling factor for which the clean system is a superfluid at lower dipolar interaction strength and a checkerboard solid at larger dipolar interaction strength. We find that, even for weak dipolar interaction, superfluidity is destroyed in favor of a Bose glass at relatively low disorder strength.  This is in contrast to what found for the Bose-Hubbard model at fixed unity filling and in the limit of weak interactions where sizable disorder strength is needed in order to destroy superfluidity. This can be explained by the hard-core nature of bosons which suppresses superfluidity even at weak dipolar interaction. Interestingly, in the presence of disorder, superfluidity persists for values of dipolar interaction strength for which the clean system is a checkerboard solid. At fixed disorder strength, as the dipolar interaction is increased, superfluidity is destroyed in favor of a Bose glass. As the interaction is further increased, the system eventually develops extended checkerboard patterns in the density distribution. Due to the presence of disorder, though, grain boundaries and defects, responsible for a finite residual compressibility, are present in the density distribution. Finally, we have studied the robustness of the superfluid phase against thermal fluctuations where we found that, at fixed dipolar interaction, the critical temperature at which superfluidity disappears decreases as the disorder strength increases.

{\textit{Acknowledgements}}  This work was supported by the NSF (PIF-1552978). The computing for this project was performed at the OU Supercomputing Center for Education and Research (OSCER) at the University of Oklahoma (OU).

\bibliography{disorder2}

\end{document}